\documentstyle[aps,psfig,prl,twocolumn]{revtex}

\title{Thermodynamically admissible form for discrete hydrodynamics}
\author{Pep Espa\~{n}ol }
\address{Dept. F\'{\i}sica Fundamental, UNED, Aptdo. 60141 E-28080, Madrid,
Spain}
\author{Hans Christian \"Ottinger}
\address{Institute of Polymers, Swiss Federal Institute of Technology, ETH-Zentrum, ML J 19, CH-8092, Z\"urich, Switzerland}

\begin{document}
\maketitle

\begin{abstract}
We construct a discrete model of fluid particles according to the
GENERIC formalism. The model has the form of Smoothed Particle
Hydrodynamics including correct thermal fluctuations. A slight
variation of the model reproduces the Dissipative Particle
Dynamics model with any desired thermodynamic behavior. The
resulting algorithm has the following properties: mass, momentum and
energy are conserved, entropy is a non-decreasing function of time and
the thermal fluctuations produce the correct Einstein distribution
function at equilibrium.
\end{abstract}

Particle based methods for solving hydrodynamic problems are good
candidates for the study of complex fluids because they allow an easy
treatment of complicated geometries as those appearing in the
interstitial regions of a colloidal or polymeric suspension. In
addition, they allow us to use molecular dynamic codes which are
comparatively much simpler than usual computational fluid dynamics
algorithms. 

Two seemingly distinct particle methods are particularly appealing
because of their versatility, Smoothed Particle Hydrodynamics (SPH)
and Dissipative Particle Dynamics (DPD).  SPH is a well-known model in
the context of computational astrophysics \cite{mon92} that is
receiving growing interest for the study of laboratory fluid dynamical
problems \cite{sph}. This method is basically a discretization of
Navier-Stokes equations on a Lagrangian grid with the aid of a weight
function. At present, there is still no version of SPH that
consistently includes thermal fluctuations, that is, there is no SPH
discretization of fluctuating hydrodynamics \cite{landau,fh}. However,
thermal fluctuations are crucial if one wants to use SPH for the study
of complex fluids. Thermal fluctuations are the ultimate responsible
for the diffusive behavior of suspended objects \cite{bedeaux}. We are
faced, therefore, with the problem of generalizing SPH to the
mesoscopic level where fluctuations are important. On the other hand,
DPD \cite{hoo92} is another particle based model that includes thermal
fluctuations in a sensible way \cite{esp95} and it has hydrodynamic
behavior \cite{hydro}. For these reasons it has been applied to a
large variety of problems dealing with complex fluids
\cite{applications}. The initial problem of non-conservation of the
energy has been resolved and the technique can now be applied to
non-isothermal problems as well \cite{energy}.  There remains,
however, a fundamental problem in DPD which is the physical
interpretation of the conservative interactions between DPD particles,
which determine the full thermodynamic behavior of the system
\cite{gro97}.

In this letter, we give a solution to these basic problems with SPH
and DPD. We formulate a Lagrangian discrete off-lattice model for
hydrodynamics which generalizes SPH by including correct thermal
fluctuations. A slight variation produces the DPD model with any
desired thermodynamic behavior.  The derivation of these models is
done within the context of the GENERIC formalism which ensures
thermodynamic consistency \cite{generic}. GENERIC postulates that any
physically sensible dynamic equation in non-equilibrium thermodynamics
has an underlying structure. All the dynamic equations studied so far
fit into the formalism. Linear irreversible thermodynamics,
non-relativistic and relativistic hydrodynamics, Boltzmann's equation,
polymer kinetic theory, and chemical reactions, just to mention a few,
have all the GENERIC structure \cite{generic,gen-applyed}. It is natural,
then, to formulate a model for fluid particles within the GENERIC
formalism. As it is becoming apparent in recent years, by putting
``more physics'' into discretization of partial differential equations
one gets improved behaviour (i.e. stability) of the discretized
equations.

The GENERIC formalism states that the time evolution equations for a 
complete set of independent variables $x$ required for the description
of a non-equilibrium system have the structure

\begin{equation}
\frac{dx}{dt}=L\!\cdot\!\frac{\partial E}{\partial x}
+M\!\cdot\!\frac{\partial S}{\partial x}.
\label{gen1}
\end{equation}
The first term in the right hand side produces the reversible part of
the dynamics whereas the second term is responsible of the
irreversible dissipative dynamics. Here, $E,S$ are the energy and
entropy of the system expressed in terms of the variables $x$ and
$L,M$ are matrices that satisfy the
following degeneracy requirements,

\begin{equation}
L\!\cdot\!\frac{\partial S}{\partial x}=0,\quad\quad\quad
M\!\cdot\!\frac{\partial E}{\partial x}=0.
\label{gen2}
\end{equation}
In addition, $L$ is antisymmetric (this guarantees that energy
is conserved) and $M$ is a positive definite symmetric matrix (this
guarantees that the entropy is a nondecreasing function of time).
If the system presents dynamical invariants $I(x)$ different from the
total energy, then further restrictions on the form of $L$ and $M$
are required
\begin{equation}
\frac{\partial I}{\partial x}\!\cdot\!
L\!\cdot\!\frac{\partial E}{\partial x}=0,\quad\quad\quad
\frac{\partial I}{\partial x}\!\cdot\!
M\!\cdot\!\frac{\partial S}{\partial x}=0.
\label{il}
\end{equation}
These conditions ensure that $d I/ dt =0$.

The deterministic equations (\ref{gen1}) are, actually, an
approximation in which thermal fluctuations are neglected. If thermal
fluctuations are not neglected, the dynamics is described by
stochastic differential equations or, equivalently, by a Fokker-Planck
equation that governs the probability distribution function
$\rho=\rho(x,t)$. This FPE has the form \cite{generic}
\begin{equation}
\partial_t\rho =
-\frac{\partial}{\partial x}\!\cdot\!
\left[\rho\left[ 
  L\!\cdot\!\frac{\partial E}{\partial x}
+ M\!\cdot\!\frac{\partial S}{\partial x} \right]
- k_B M\!\cdot\!\frac{\partial \rho}{\partial x}\right],
\label{FPE}
\end{equation}
where $k_B$ is Boltzmann's constant. We require that the equilibrium
solution of this Fokker-Planck equation is given by the Einstein
distribution function in the presence of dynamical invariants
\cite{mixing}, this is
\begin{equation}
\rho^{\rm eq}(x) = g(E(x),I(x))\exp \{ S(x)/k_B\},
\label{einst}
\end{equation}
where the function $g$ is completely determined by the arbitrary
initial distribution of dynamical invariants. This imposes further
conditions on the matrices $L,M$, namely,

\begin{equation}
\frac{\partial}{\partial x}\!\cdot\!\left[
L \!\cdot\!\frac{\partial E}{\partial x}\right]=0,\quad\quad\quad
M\!\cdot\!\frac{\partial I}{\partial x}=0.
\label{add}
\end{equation}
The first property can be derived independently with projection
operator techniques \cite{generic}. The second property implies that the
last equation in (\ref{il}) is automatically satisfied. When
fluctuations are present, the entropy functional ${\cal S}[\rho]=\int
S(x)\rho(x,t)dx -k_B\int \rho(x,t)\ln \rho(x,t) dx$ plays the role of
a Lyapunov function with $\partial_t {\cal S}[\rho]\ge 0$.

We finally consider the It\^o stochastic differential equations that
are mathematically equivalent to the above Fokker-Planck equation
\cite{stoch}

\begin{equation}
dx = \left[
L\!\cdot\!\frac{\partial E}{\partial x}
+M\!\cdot\!\frac{\partial S}{\partial x}
+k_B\frac{\partial }{\partial x}\!\cdot\!M\right]dt
+d\tilde{x}.
\label{sde1}
\end{equation}
Here, the stochastic term $d\tilde{x}$ is a linear combination of
independent increments of the Wiener process. It satisfies the
mnemotechnical It\^o rule

\begin{equation}
d\tilde{x}d\tilde{x}^T=2k_BM dt,
\label{F-D}
\end{equation}
which means that $d\tilde{x}$ is an infinitesimal of order $1/2$
\cite{stoch}.  Eqn. (\ref{F-D}) is a compact and formal statement of
the fluctuation-dissipation theorem.

When formulating new models it might be convenient to specify
$d\tilde{x}$ directly instead of $M$. This ensures that $M$ through
(\ref{F-D}) automatically satisfies the symmetry and positive definite
character. In order to guarantee that the total energy and dynamical
invariants do not change in time, a strong requirement on the form of
$d\tilde{x}$ holds,
\begin{equation}
\frac{\partial E}{\partial x}\!\cdot\! d\tilde{x}=0,\quad\quad
\frac{\partial I}{\partial x}\!\cdot\! d\tilde{x}=0,
\label{cons}
\end{equation}
implying the last equations in (\ref{gen2}) and (\ref{add}).

The basic problem in any non-equilibrium description of a given system
is to identify the relevant set of variables in the system. We aim at
modeling {\em fluid particles}, that is, portions of fluid or large
clusters of molecules that move following collective motions. It is
sensible to assume that these fluid particles are small moving
thermodynamic systems with the center of mass located at ${\bf r}_i$,
and with momentum ${\bf p}_i$, volume ${\cal V}_i$, mass $m_i$, and
entropy $s_i$ (or internal energy $\epsilon_i$).  Even though they are
thermodynamic systems in the sense that an entropy function can be
defined, the fluid particles are assumed to be small enough to suffer
from stochastic fluctuations due to the underlying molecules forming
the fluid particle. The state of the system is $x=\{{\bf r}_i,
{\bf p}_i,{\cal V}_i,m_i,s_i,\;\;i=1,\ldots,N\}$, where $N$ is the
number of fluid particles.

The two basic building blocks in GENERIC are the energy and the
entropy of the system as a function of the selected variables. In our
case they are

\begin{equation}
E(x)=\sum_i \frac{{\bf p}^2_i}{2m_i}
+\epsilon({\cal V}_i,m_i,s_i),\quad\quad S(x) = \sum_i s_i,
\label{ES}
\end{equation}
where $\epsilon({\cal V}_i,m_i,s_i)$ is the internal energy as a
function of the extensive variables of the fluid particle. We will
assume the hypothesis of local equilibrium, in accordance with the
usual treatment of hydrodynamics. Therefore, the internal energy of
the fluid particles is the same function of the mass, volume, and
entropy as the total internal energy of the whole fluid system at equilibrium.
We can now consider the derivatives of $E$ and $S$ which are given
by

\begin{equation}
\frac{\partial E}{\partial x} =
\left(
\begin{array}{c} {\bf 0}\\{\bf v}_i\\-p_i\\\mu_i-\frac{1}{2}v_i^2\\T_i
\end{array}\right),
\quad\quad\quad
\frac{\partial S}{\partial x} =
\left(
\begin{array}{c} {\bf 0}\\ {\bf 0}\\ 0\\ 0\\ 1
\end{array}\right),
\label{derES}
\end{equation}
where ${\bf v}_i= {\bf p}_i/m_i$ is the velocity of the fluid
particles and the intensive parameters are the pressure $p_i$, the
chemical potential per unit mass $\mu_i$, and the temperature $T_i$
which are functions of the extensive variables ${\cal V}_i, m_i,s_i$
of the fluid particles. We have to consider also the dynamical
invariants $I(x)$ that we want to retain in the discrete model. These
dynamical invariants are the total momentum ${\cal {\bf P}}=\sum_i
{\bf p}_i$, the total volume ${\cal V}= \sum_i {\cal V}_i$, and the
total mass ${\cal M}= \sum_i m_i$.

We construct now the $L$ matrix for the discrete hydrodynamics
problem.  We impose that the reversible part of the equations of
motion should give $\dot{\bf r}_i|_{\rm rev} = {\bf v}_i$,
$\dot{m}_i|_{\rm rev} = \dot{s}_i|_{\rm rev}= 0$. That is, the
position of the fluid particles changes according to its velocity, the
``identity'' of the fluid particles is given by the constant mass it
possesses, and finally, the reversible part of the dynamics should not
produce any change in the entropy content of the fluid particle. By
looking at the term $L\!\cdot\!\partial E/\partial x$ in
Eqn. (\ref{sde1}) with the energy derivatives given in (\ref{derES}),
the $L$ matrix that produces the desired equations can only be of the
form of $N\times N$  blocks
${\bf L}_{ij}$ of size $9\times 9$ of the form

\begin{equation}
{\bf L}_{ij}=
\left(
\begin{array}{ccccccccc}
 {\bf 0} && {\bf 1}\delta_{ij} && {\bf 0} &&{\bf 0} &&{\bf 0} \\
-{\bf 1}\delta_{ij} && {\bf 0} && {\bf \Omega}_{ij} &&{\bf 0} &&{\bf 0} \\
{\bf 0} &&-{\bf \Omega}_{ji}^T&&  0 &&  0 && 0 \\
{\bf 0} && {\bf 0} && 0 && 0 && 0 \\
{\bf 0} && {\bf 0} && 0 && 0 && 0 \\
\end{array}\right),
\label{L}
\end{equation}
The two last rows ensure the invariance of $m_i,s_i$, the two last
columns are fixed by antisymmetry. The first row ensures the equation
of motion for the position and the first column is fixed by
antisymmetry. We still have freedom for the form of the vector ${\bf
\Omega}_{ij}$ which can, in principle, depend on all state variables.
$L$ is antisymmetric because ${\bf L}_{ij}=-{\bf
L}^T_{ji}$ and it satisfies Eqn. (\ref{gen2}). The
effect on the invariants, Eqn. (\ref{il}), is guaranteed if ${\bf
\Omega}_{ij}$ obeys
\begin{equation}
\sum_i{\bf \Omega}_{ij}=\sum_i{\bf \Omega}_{ji}=0.
\label{o0}
\end{equation}
Finally, we can guarantee  the first property in Eqn. (\ref{add}) if
the following identity is satisfied
\begin{equation}
\sum_{ij}\left[
\frac{\partial}{\partial {\bf p}_i}
\!\cdot\!{\bf \Omega}_{ij}p_j
+
\frac{\partial}{\partial {\cal V}_i}
{\bf \Omega}_{ji}\!\cdot\!{\bf v}_j\right]
=0.
\label{cond1}
\end{equation}
The resulting reversible equations for momentum and volume are
\begin{equation}
\dot{\bf p}_i|_{\rm rev} = -\sum_j{\bf \Omega}_{ij}p_j,\quad\quad\quad
\dot{\cal V}_i|_{\rm rev} = -\sum_j{\bf \Omega}_{ji}\!\cdot\!{\bf v}_j.
\label{rev1}
\end{equation}
Note that the vector ${\bf \Omega}_{ij}$ can be interpreted as a discrete
version of the gradient operator in such a way that the above equations
look like a discrete version of the momentum equation and continuity
equation (in terms of the specific volume $1/\rho$ instead of the mass
density $\rho$) of a non-dissipative fluid in a Lagrangian description.
We propose the following form for ${\bf \Omega}_{ij}$

\begin{equation}
{\bf \Omega}_{ij} = \mbox{\boldmath $\omega$}_{ij}-\frac{1}{N}\left[
\sum_k\mbox{\boldmath $\omega$}_{ik}-\sum_k\mbox{\boldmath $\omega$}_{jk}\right]
\label{omega}
\end{equation}
where $\mbox{\boldmath $\omega$}_{ij}
=({\cal V}/N)^2\mbox{\boldmath $\nabla$} \Delta(r_{ij})$, and
$\Delta(r)$ being a weight function of range $r_c$ normalized to
unity, $\int d{\bf r}\Delta(r)=1$. This form (\ref{omega}) satisfies
Eqns. (\ref{o0}) and (\ref{cond1}). If the range $r_c$ is much smaller
than the typical length scale of variation of the hydrodynamic variables
{\em and} the typical distance between points is much smaller
than $r_c$, then it is possible to prove that
\begin{equation}
{\cal V}_i\mbox{\boldmath $\nabla$} f(r_i)\sim\sum_j\mbox{\boldmath $\omega$}_{ij}f(r_j)
\end{equation}
Under these circumstances, the discrete equations (\ref{rev1}) converge
towards the continuum Euler equations of hydrodynamics.

Our aim now is to construct the matrix $M$. We have to specify first
where irreversibility occurs. We do not want irreversible processes
associated with the evolution of the position, volume or mass of the
particles.  This implies that the noise term in the
equation of motion (\ref{sde1}) has the structure $d\tilde{x}^T\rightarrow
\left( {\bf 0},d\tilde{\bf p}_i,0,0,d\tilde{s}_i\right)$. 

Thermal fluctuations are introduced into the equations of
hydrodynamics through a random stress tensor and random heat flux
\cite{landau}.  By simple analogy with the expression of the random
noise in non-linear hydrodynamics \cite{esp97}, we postulate the
following random terms

\begin{eqnarray}
d\tilde{\bf p}_i &=& \sum_j{\bf \Omega}_{ij}\!\cdot\!d\tilde{\bf \sigma}_j,
\nonumber\\
d\tilde{s}_i &=& \frac{1}{T_i}\sum_j{\bf \Omega}_{ij}\!\cdot\!d\tilde{\bf J}^q_j
+\frac{1}{T_i}d\tilde{\bf \sigma}_i:\sum_j{\bf \Omega}_{ij}{\bf v}_j^T,
\label{ran2}
\end{eqnarray}
where the random stress $d\tilde{\bf \sigma}_i $ and random heat 
flux $d\tilde{\bf J}^q_i $ are defined by

\begin{eqnarray}
d\tilde{\bf \sigma}_i &=&
(4k_BT_i\eta_i)^{1/2}\overline{d{\bf W}}^{S}_{i}
+(6k_BT_i\zeta_i)^{1/2}{\bf 1}\frac{1}{D}{\rm tr}[d{\bf W}_{i}],
\nonumber\\
d\tilde{\bf J}^q_i &=& T_i(2k_B\kappa_i)^{1/2}d{\bf V}_i.
\label{ran3}
\end{eqnarray}
Here, $\eta_i$ is the shear viscosity, $\zeta_i$ is the bulk viscosity,
and $\kappa_i$ is the
thermal conductivity. The traceless symmetric
random matrix $\overline{d{\bf W}}^{S}_i$ is given by
\begin{equation}
\overline{d{\bf W}}^S_{i}=
\frac{1}{2}\left[d{\bf W}_{i}+d{\bf W}^T_{i}\right]
-\frac{1}{D}{\rm tr}[d{\bf W}_{i}]{\bf 1}.
\label{decomp}
\end{equation}
$D$ is the physical dimension of space and $d{\bf W}_{i}$ is a matrix
of independent Wiener increments.  The vector $d{\bf V}_i$ is also a
vector of independent Wiener increments.  They satisfy the It\^o
mnemotechnical rules
\begin{eqnarray}
d{\bf W}^{\mu\mu'}_{i}d{\bf W}^{\nu\nu'}_{j}&=&
\delta_{ij}\delta_{\mu\nu}\delta_{\mu'\nu'}dt,
\nonumber\\
d{\bf V}^{\mu}_id{\bf V}^{\nu}_j&=&\delta_{ij}\delta_{\mu\nu}dt,
\nonumber\\
d{\bf V}^{\mu}_id{\bf W}^{\nu\nu'}_j&=&0.
\label{ran3b}
\end{eqnarray}
Note that the postulated forms for $d\tilde{{\bf p}}_i,d\tilde{s}_i$ in
Eqn. (\ref{ran2}) satisfy $\sum_i{\bf
v}_i\!\cdot\!d\tilde{{\bf p}}_i+T_id\tilde{s}_i=0$ and
$\sum_id\tilde{{\bf p}}_i=0$ and, therefore, Eqns. (\ref{cons}) are
satisfied. This means that the postulated noise terms conserve
momentum and energy exactly. It is now a matter of algebra to
construct the dyadic $d\tilde{x}d\tilde{x}^T$ and from
Eqn. (\ref{F-D}) extract the matrix $M$. Once $M$ is constructed, the
final equations of motion for the discrete hydrodynamic variables are
($D=3$ and, for simplicity, constant transport coefficients are
assumed),

\begin{eqnarray}
d{m}_i &=& 0,\quad\quad\quad d{\bf r}_i = {\bf v}_i dt,
\quad\quad \quad
d{\cal V}_i = D_i dt
\nonumber\\
d{\bf p}_i &=& -\left(\sum_j{\bf \Omega}_{ij} p_j 
+\sum_j{\bf \Omega}_{ij} \!\cdot\!{\bf \sigma}_j\right)dt + d\tilde{\bf p}_i
\nonumber\\
T_id{s}_i &=&\left(1-\frac{k_B}{C_{Vi}}\right)
\left(2\eta\overline{\bf G}_i:\overline{\bf G}_i+\zeta D_i^2\right)dt 
\nonumber\\
&-&\sum_j{\bf \Omega}_{ij} \!\cdot\!{\bf J}^q_jdt +T_id\tilde{s}_i
\label{eqmot}
\end{eqnarray}
We have introduced the stress tensor ${\bf \sigma}_i$, the traceless
symmetric velocity ``gradient'' tensor $\overline{\bf G}_i$, and the
``divergence'' $D_i$ by

\begin{eqnarray}
{\bf \sigma}_i^{\mu\nu} &=&-2\eta\overline{\bf G}_i^{\mu\nu}
-\zeta D_i\delta^{\mu\nu}
\nonumber\\
\overline{\bf G}_i^{\mu\nu} &= &
\frac{1}{2}\sum_k[
{\bf \Omega}_{ik}^{\mu}{\bf v}_k^\nu+
{\bf \Omega}_{ik}^{\nu}{\bf v}_k^\mu]-
\frac{1}{3}\delta^{\mu\nu}\sum_k{\bf \Omega}_{ik}\!\cdot\!{\bf v}_k
\nonumber\\
 D_i &=&\sum_k{\bf \Omega}_{ik}\!\cdot\!{\bf v}_k
\label{disc}
\end{eqnarray}
The heat flux ${\bf J}^q_i$ of particle $i$ is defined
by,

\begin{equation}
{\bf J}^q_i=T_i^2\kappa\sum_k{\bf \Omega}_{ik}\left[
\frac{1}{T_k}+\frac{k_B}{C_{Vk}}(1+\delta_{ik})\frac{1}{T_k}\right]
\label{fluxes}
\end{equation}
One can easily recognize all the terms corresponding to the continuum
equations of hydrodynamics \cite{de-groot}. The factor $k_B/C_{Vi}$,
where $C_{Vi}$ is the specific heat at constant volume of particle
$i$, comes from the term $k_B\partial\!\cdot\!M/\partial x$ in
Eqn. (\ref{sde1}). In the continuum version of hydrodynamics, this
term is zero due to locality \cite{saa82}.

Eqns. (\ref{eqmot}) are the main result of this Letter.  They have the
structure of SPH but conserve total mass, total momentum, total energy
and total volume of the particles. Because of the GENERIC structure of
the equations, the entropy functional ${\cal S}[\rho]$ of the system
is a non-decreasing function of time and the equilibrium solution is
given by Einstein distribution function.

Instead of the noise terms (\ref{ran2}) one could also postulate
the noise terms as in DPD \cite{energy}

\begin{eqnarray}
d\tilde{\bf p}_i&=&\sum_j {\bf B}_{ij} dW_{ij}
\nonumber\\
d\tilde{s}_i &=& -\frac{1}{T_i}\sum_j {\bf B}_{ij}\!\cdot\!{\bf v}_idW_{ij}
+\frac{1}{T_i}\sum_jA_{ij}dV_{ij}
\end{eqnarray}
where ${\bf B}_{ij}=-{\bf B}_{ji}$ and $A_{ij}=A_{ji}$ are suitable
functions of position and, perhaps, other state variables. 
The independent Wiener processes satisfy $dW_{ij}=dW_{ji}$ and
$dV_{ij}=-dV_{ji}$ and the following It\^o mnemotechnical rules
\begin{eqnarray}
dW_{ii'}dW_{jj'} &=& [\delta_{ij}\delta_{i'j'}+\delta_{ij'}\delta_{i'j}]dt
\nonumber\\
dV_{ii'}dV_{jj'} &=& [\delta_{ij}\delta_{i'j'}-\delta_{ij'}\delta_{i'j}]dt
\label{wie}
\end{eqnarray}
These noise terms satisfy also Eqn. (\ref{cons}) and momentum and
energy are conserved. The resulting dynamic equations have the same
reversible part as in Eqn. (\ref{eqmot}) and have the familiar
``Brownian dashpot'' dissipative forces of the DPD model. In this way,
by introducing a volume and an internal energy variables into the
standard DPD model, one derives a thermodynamically consistent model
in which the ``conservative'' forces are truly pressure forces between
DPD particles. The resulting DPD equations are simpler than
Eqns. (\ref{eqmot}) and produce macroscopic hydrodynamic behaviour,
but the identification of the actual values of the transport
coefficients is less obvious.

P.E. wishes to acknowledge useful conversations with M. Serrano and 
M. Ripoll and finantial support from DGYCIT Project No PB97-0077.

\end{document}